\documentclass[final,numberedheadings]{aipproc}
\usepackage{amsmath}
\usepackage{amsfonts}
\usepackage{amssymb}
\usepackage{MHequ}
\layoutstyle{6x9}

\def\D{{d_{\rm int}}}

\def\H{{\rm H}}
\def\C{{\rm C}}
\def\L{{\rm L}}
\def\R{{\rm R}}

\def\emu{\bar{\mu}}
\def\kB{k_{\rm B}}

\def\nablax{~\partial_x}
\def\eref#1{(\ref{#1})}
\let\rho=\varrho

\def\dof{{{\em d.o.f.}}}

\def\dT{\nablax\left(\frac{1}{T}\right)}

\def\dmuu{\nablax\left(-\frac{\mu}{T}\right)}

\def\B{\frac{\lambda p_t(l) l}{(2\pi  m)^{1/2}}}

\begin{document}

\title{Thermoelectric transport in billiard systems}


\author{Giulio Casati}{
  address={Center for  Nonlinear and  Complex  Systems, Universit\`a
  degli Studi dell'Insubria, Como Italy}
,altaddress={CNR-INFM and Istituto Nazionale di Fisica Nucleare, Sezione di Milano}
}
\author{Carlos Mej\'ia-Monasterio}{
  address={Istituto dei Sistemi Complessi, Consiglio Nazionale delle Ricerche, Sesto Fiorentino Italy}
}

\begin{abstract}
We discuss the thermoelectric (TE) transport in billiard systems of
interacting particles, coupled to stochastic particle reservoirs.  Recently in
\cite{CMMP08}, analytical exact expressions for the TE transport of
noninteracting gases of polyatomic molecules were obtained, and a novel
microscopic mechanism for the increase of thermoelectric efficiency described.
After briefly reviewing the derivation of \cite{CMMP08}, in this paper we
focus on the effects that the particle-particle interaction has on the TE
efficiency.  We show that interaction reduces the maximal thermodynamic
efficiency.  However, the mechanism for the efficiency's increase towards its
Carnot upper limit, described in \cite{CMMP08}, remains unaffected.
\end{abstract}

\maketitle

\section{Introduction}

Thermoelectricity concerns the conversion of temperature differences into
electric potential or vice-versa.  It can be used to perform useful electrical
work or to pump heat from cold to hot place, thus performing refrigeration.
Although thermoelectricity was discovered almost 200 years ago, a strong
interest of the scientific community arose only in the 1950's when Abram Ioffe
discovered that doped semiconductors exhibit relatively large thermoelectric
effect. This initiated an intense research activity in semiconductors physics
which was not motivated by microelectronics but by the Ioffe suggestion that
home refrigerators could be built with semiconductors \cite{mahan,majumdar}.
As a result of these efforts the thermoelectric material
$\textrm{Bi}_2\textrm{Te}_3$ was developed for commercial purposes.  However
this activity lasted only few years until the mid 1960's since, in spite of
all efforts and consideration of all type of semiconductors, it turned out
that thermoelectric refrigerators have still poor efficiency as compared to
compressor based refrigerators.  Nowadays Peltier refrigerators are mainly
used in situations in which reliability and quiet operation, and not the cost
and conversion efficiency, is the main concern, like equipments in medical
applications, space probes etc.  In the last decade there has been an
increasing pressure to find better thermoelectric materials with higher
efficiency.  The reason is the strong environmental concern about
chlorofluorocarbons used in most compressor-based refrigerators.  Also the
possibility to generate electric power from waste heat using thermoelectric
effect is becoming more and more interesting
\cite{dresselhaus,mahan,majumdar}.
 
The suitability of a thermoelectric material for energy conversion or
electronic refrigeration is evaluated by the thermoelectric figure of merit
$Z$,
\begin{equ} \label{eq:ZT-def}
Z = \frac{\sigma S^2}{\kappa} \ ,
\end{equ}
where $\sigma$ is the coefficient of electric conductivity, $S$ is the Seebeck
coefficient and $\kappa$ is the thermal conductivity. The Seebeck coefficient
$S$, also called thermopower, is a measure of the magnitude of an induced
thermoelectric voltage in response to a temperature difference across the
material.

For a given material, and a pair of temperatures $T_\H$ and $T_\C$ of hot and
cold thermal baths respectively, $Z$ is related to the {\em efficiency} $\eta$
of converting the heat current $J_Q$ (between the baths) into the electric
power $P$ which is generated by attaching a thermoelectric element to an
optimal Ohmic impedance. Namely, in the linear regime:
\begin{equ} \label{eq:efficiency}
\eta = \frac{P}{J_Q} = \eta_\mathrm{carnot} \cdot \frac{\sqrt{ZT + 1} -
  1}{\sqrt{ZT + 1} + 1} \ ,
\end{equ}
where $\eta_\mathrm{carnot}=1-T_\C/T_\H$ is the Carnot efficiency and $T =
(T_\H + T_\C)/2$. Thus a good thermoelectric device is characterized by a
large value of the non-dimensional figure of merit $ZT$.

Since the 1960's many materials have been investigated but the maximum value
found for $ZT$ was achieved for the
$(\textrm{Bi}_{1-x}\textrm{Sb}_x)_2(\textrm{Se}_{1-y}\textrm{Te}_y)_3$ alloy
family with $ZT \approx 1$.  However, values $ZT > 3$ are considered to be
essential for thermoelectrics to compete in efficiency with mechanical power
generation and refrigeration at room temperatures.  The efforts recently
focused on a bulk of new advanced thermoelectric materials and on
low-dimensional materials, and only a small increment of the efficiency, $ZT
\lesssim 2.6$, has been obtained \cite{dresselhaus}.

One of the main reasons for this partial success is a limited understanding of
the possible microscopic mechanisms leading to the increase of $ZT$, with few
exceptions \cite{linke}. From a dynamical point of view, cross effects in
transport have been barely studied \cite{microscopic,MMLL}.  So far, the
challenge lies in engineering a material for which the values of $S$, $\sigma$
and $\kappa$ can be controlled independently.  However, the different
transport coefficients are interdependent, making optimization extremely
difficult.

In a recent paper, we have studied the thermoelectric process in a gas of
non-interacting polyatomic molecules \cite{CMMP08}.  We showed that large
values of $ZT$, in principle approaching to Carnot's efficiency, are obtained
when the number of the molecule's internal degrees of freedom (\dof) is
increased. Using the rotating-disk interaction introduced in \cite{MMLL}, in
this paper, we study the effect that a generic particle-particle interaction
has on the TE efficiency of these systems.  In the following section
\ref{sec:TE}, we review briefly the microscopic expressions for the TE process
in ergodic ideal gases obtained in \cite{CMMP08}.  In the section
\ref{sec:int}, we study numerically, the modifications to these expressions,
due to the interaction. Our conclusions appear in section \ref{sec:concl}.

\section{Thermoelectric Transport}
\label{sec:TE}

In the linear response regime (see {\em e.g.}  \cite{bergman}), the transport
equations for a thermoelectric material give the heat current $J_Q$ and the
electric current $J_e$ through an homogeneous sample subjected to a
temperature gradient $\nablax T$ and a electrochemical potential gradient
$\nablax \emu$ as
\begin{equa}[2] \label{eq:Ju1}
& J_Q & \ = \ & - \kappa' \nablax T - T \sigma S \nablax\emu\ ,
\\
& J_e & \ = \ & - \sigma S \nablax T - \sigma \nablax\emu\ .
\end{equa}
Here and in what follows, we assume that the transport occurs along the
$x$-direction and the temperature is given in units where the Boltzmann
constant $\kB=1$.

The electrochemical potential is the sum of a chemical and an electric part
$\emu = \mu + \mu_e$, where $\mu$ is the chemical potential of the particles
and, if $e$ is the particle's charge, $\mu_e = e\phi$ is the work done by the
particles against an external electric field ${\cal E}=-\nablax\phi$.  From
(\ref{eq:Ju1}) the usual phenomenological relations follow: if the thermal
gradient vanishes, $\nablax T=0$, then $J_e = -\sigma \nablax \phi = \sigma
{\cal E}$, since for an isothermal homogeneous system $\mu$ is uniform.  If
the electric current vanishes, $J_e = 0$, then $\nablax\emu = S \nablax T$,
which is the definition of the Seebeck coefficient, and $J_Q = -\kappa \nablax
T$ where $\kappa = \kappa' - T \sigma S^2$ is the usual thermal conductivity
(see {\em e.g.} \cite{domenicali}).

From the theory of irreversible thermodynamics, $\mu$ and $\mu_e$ cannot be
determined separately; only their combination in $\emu$ appears in
\eref{eq:Ju1} \cite{walstrom}.  Based on this equivalence, in what follows we
take into account the chemical part only, {\em i.e.}, $\bar{\mu} = \mu$.

Since in the linear regime in which dissipative effects, such as, {\em e.g.},
Joule heating, can be neglected and at low particle densities, thermodiffusion
and thermoelectricity are equivalent process, we can study TE in terms of the
thermodiffusion transport equations
\begin{equa}[2] \label{eq:Ju2}
& J_u & \ = \ & L_{uu}\dT + L_{u\rho}\dmuu , \\ 
& J_\rho& \ = \ & L_{\rho u}\dT + L_{\rho\rho}\dmuu\ ,
\end{equa}
where ${J}_u$ and $J_\rho = J_e/e$ are the energy and particle density
currents, and $\L_{u\rho} = L_{\rho u}$ follows from the Onsager reciprocity
relations.  From the entropy balance equation for open systems \cite{callen}
\begin{equ} \label{eq:Ju-def}
J_u = J_Q + \mu J_\rho \ ,
\end{equ}
and substituting $J_Q$ in \eref{eq:Ju1} in favor of $J_u$ and comparing the
resulting equations with \eref{eq:Ju2} it follows that the TE transport
coefficients can be written in terms of the thermodiffusion $L$-coefficients
as
\begin{equ}\label{eq:coefs}
\sigma  =  \frac{e^2}{T}L_{\rho\rho} \ , \quad
\kappa  =  \frac{1}{T^2}\frac{\det \mathbb{L}}{L_{\rho \rho}} \ , \quad
S  =  \frac{1}{eT}\left(\frac{L_{u\rho}}{L_{\rho\rho}} - \mu\right) \ .
\end{equ}
Furthermore, from Eq.  \eref{eq:ZT-def}, we obtain for the figure of merit
\begin{equ}\label{eq:ZT}
ZT = \frac{\left(L_{u\rho} - \mu L_{\rho\rho}\right)^2}{\det \mathbb{L}}\ .
\end{equ}
Note that in Eqs.~\eref{eq:coefs} and \eref{eq:ZT}, $T$ and $\mu$ are taken as
the mean values.

\begin{figure}
  \includegraphics[height=.2\textheight]{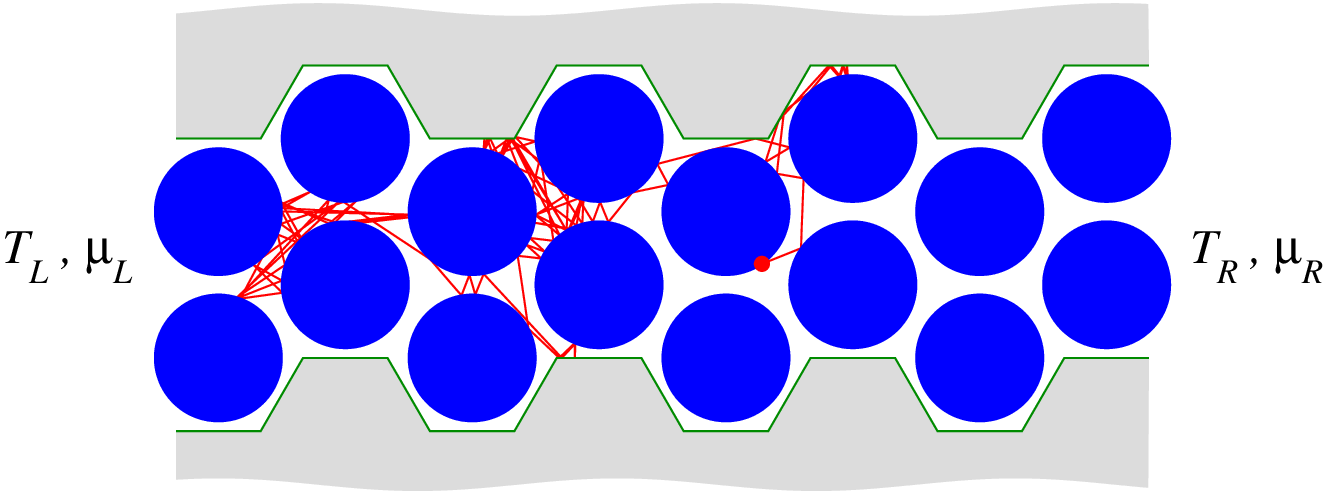}
  \caption{The open Lorentz gas system and a typical particle's trajectory.
The particles are scattered from fixed disks of radius $R$ disposed in a
triangular lattice at critical horizon, {\em i.e.}, the width and height of
the cells are $\Delta x = 2R$ and $\Delta y = 2W$ respectively, where
$W=4R/\sqrt{3}$ is the separation between the centers of the disks.  The
channel is coupled at the left and right boundaries to two thermochemical
baths at temperatures $T_\L$ and $T_\R$ and chemical potentials $\mu_\L$ and
$\mu_\R$, respectively.\label{fig:billiard}}
\end{figure}

Consider now an ergodic gas of non-interacting, electrically neutral particles
of mass $m$ with $d_{\rm int}$ internal \dof ~(rotational or vibrational),
enclosed in a $d$ dimensional container.  To study the non-equilibrium state
of such dilute poly-atomic gas we consider a chaotic billiard channel (like
the one shown in Fig.~\ref{fig:billiard}) connected through openings of length
$\lambda$ to two {\em reservoirs} of particles which are idealized as infinite
chambers with the same poly-atomic gas at equilibrium density $\rho$ and
temperature $T$.  From the reservoirs, particles are injected into the channel
at a rate $\gamma$, as explained in the appendix \ref{app:baths}

The particle injection rate $\gamma$ is related to the value of the chemical
potential $\mu$ at the reservoirs which, for a gas of polyatomic molecules
with a total of $D=d + d_{\rm int}$ \dof, at equilibrium density $\rho$ and
temperature $T$ reads
\begin{equ} \label{eq:chempotMM}
\mu = \mu_0 + T \ln \left(\frac{\rho}{T^{D/2}}\right) = \mu_0' + T \ln\left(\frac{\gamma}{T^{(D+1)/2}}\right) \ ,
\end{equ}
where $\mu_0$ and $\mu_0'$ are reference values of the chemical potential and
the second equality is simply obtained after substitution of $\rho$ from
Eq.~\eref{eq:gamma}.  Furthermore, energy is injected from the reservoirs at a
rate $\varepsilon = \gamma T(D+1)/2$ (see appendix \ref{app:baths}).

Calling $p_t(l)$ the transmission probability of the channel of length $l$,
the density currents $J_u, J_\rho$ for noninteracting particles
\cite{phonons}, assume a simple form: they are $p_t(l)$ times the difference
between the left and right corresponding injection rates,
$\varepsilon,\gamma$, respectively, namely
\begin{equ} \label{eq:noint-currents-1}
J_\rho \ = \  p_t\left(\gamma_\L - \gamma_\R\right)\ , \quad \quad
J_u    \ = \  p_t\left(\varepsilon_\L -  \varepsilon_\R\right)\ .
\end{equ}
Using \eref{eq:chempotMM} to eliminate $\gamma$ in favor of $\mu$ we obtain,
\begin{equa}[2] \label{eq:noint-currents-2}
&J_\rho &\ = \ &  -\B ~\nablax\left(T^{(D+1)/2} \ e^{\ \mu/T}\right) \
,\\  
&J_u & \ = \ & -\B ~\frac{D+1}{2}\nablax\left(T^{(D+3)/2} \ e^{\ \mu/T}\right)\ ,
\end{equa}
Taking total differentials of \eref{eq:noint-currents-2} in the variables
$1/T$ and $\mu/T$ and comparing the resulting expression with
Eq.~\eref{eq:Ju2} we obtain exact microscopic expressions for the Onsager
coefficients and thus, for the TE transport coefficients, namely
\def\G{\frac{\lambda p_t l}{(2\pi m)^{1/2}}}
\begin{equa}[2] \label{eq:micro-TE}
& \sigma & \ = \ & \G\frac{e^2\rho}{T^{1/2}} \ , \\
& S & \ = \ & \frac{1}{e}\left(\frac{D+1}{2}\right) \ , \\
& \kappa & \ = \ & \G \left(\frac{D+1}{2}\right) \rho T^{1/2} \ .
\end{equa}
Note that for a chaotic billiard channel with a diffusive dynamics, the
transmission probability decays as $p_t(l) \propto l^{-1}$ which means that
all the elements of the Onsager matrix $\mathbb{L}$ become size independent.

Finally, plugging \eref{eq:micro-TE} into \eref{eq:ZT-def} and noting that
$c^*_V=D/2$ is the dimensionless heat capacity at constant volume of the gas,
we obtain
\begin{equ}\label{eq:ZTmain}
ZT = \frac{1}{\hat{c}_V}\left(\hat{c}_V - \frac{\mu}{T}\right)^2 \ ,
\end{equ}
where for simplicity we have called $\hat{c}_V = c^*_V + 1/2$. A particular
case of \eref{eq:ZTmain} was previously obtained, for noninteracting
monoatomic ideal gases in $3$ dimensions \cite{vining}.

\begin{figure}[!t]
  \includegraphics[height=.4\textheight]{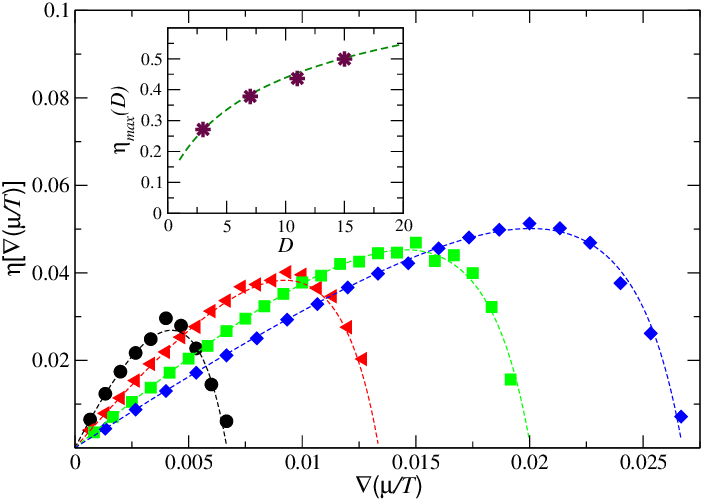}
  \caption{Thermodynamic    efficiency    $\eta$     as    a    function    of
  $\nabla\left(\mu/T\right)$ for  a chain of $15$  cells, $T_L=95$, $T_R=105$,
  and  $D=3$  (circles),  $D=7$   (squares),  $D=11$  (diamonds),  and  $D=15$
  (triangles).   The  dashed  lines  correspond  to  the  analytical  solution
  obtained      from     substituting      \eref{eq:noint-currents-2}     into
  \eref{eq:efficiency-2}. In  the inset,  maximum efficiency obtained  for the
  different values  of $D$ (crosses), in units  of the $\eta_\mathrm{carnot}$.
  The        dashed        line        corresponds        to        expression
  \eref{eq:efficiency}.\label{fig:eta}}
\end{figure}

The analytical expressions for the TE transport \eref{eq:micro-TE} and
\eref{eq:ZTmain} are exact \cite{CMMP08}. They predict and increase of the TE
efficiency with $D$. To verify this and further compare in the next section
with the interacting particles, we have considered a gas of composite
particles with $\D \ge 1$ internal rotational \dof \cite{CMMP08} inside a
Lorentz gas channel (Fig.~\ref{fig:billiard}) coupled to two thermochemical
baths. The model of composite particles and its dynamics is explained in the
appendix \ref{app:colrules}.  We have numerically measured the TE efficiency
$\eta$ as the ratio
\begin{equ} \label{eq:efficiency-2}
\eta = \frac{e J_\rho \Delta\mu}{J_u} \ ,
\end{equ}
where $\Delta\mu = \mu_\R -  \mu_\L$ is the chemical potential difference. For
a  fixed temperature  gradient,  in  Fig.~\ref{fig:eta} we  show  $\eta$ as  a
function of  $\nabla(\mu/T)$ for different  values of $D$. Note  the excellent
agreement between the numerical data  (symbols) and the analytical solution of
$\eta$,    obtained   from   substituting    \eref{eq:noint-currents-2}   into
\eref{eq:efficiency-2}. In  each case, the efficiency reaches  a maximum value
$\eta_{max}$ for  some optimal  value of $\nabla(\mu/T)$.   Moreover, $\eta=0$
occurs  when  $J_\rho=0$,  which  in   turns  is  determined  by  the  Seebeck
coefficient.    Consistently  with   Eq.~\eref{eq:micro-TE},   the  value   of
$\nabla(\mu/T)$ for  which $\eta=0$ grows  linearly with $D$. Finally,  in the
inset of Fig.~\ref{fig:eta} we show the numerically obtained $\eta_{max}$ as a
function of  $D$ (stars)  and compare them  with the analytical  expression of
\eref{eq:efficiency}, with an excellent agreement.

\section{Particle-particle interaction}
\label{sec:int}

We now turn our attention to the interacting case. In absence of particles'
interaction, $ZT$ is independent of the sample size $l$ and depends on the
temperature only through the chemical potential term.  This is due to the fact
that with no interactions, $p_t$ depends on the geometry of the billiard only.
From a physical point of view this means that the mean free path of the gas
particles is energy independent. When the particles interacting, in general
$p_t$ depends on the local density and temperature of the gas, leading to a
more realistic situation \cite{EMMZ}. Therefore, the analytical expressions
for the TE transport coefficients \eref{eq:micro-TE} are not longer exact.

\begin{figure}[!t]
  \includegraphics[scale=0.95]{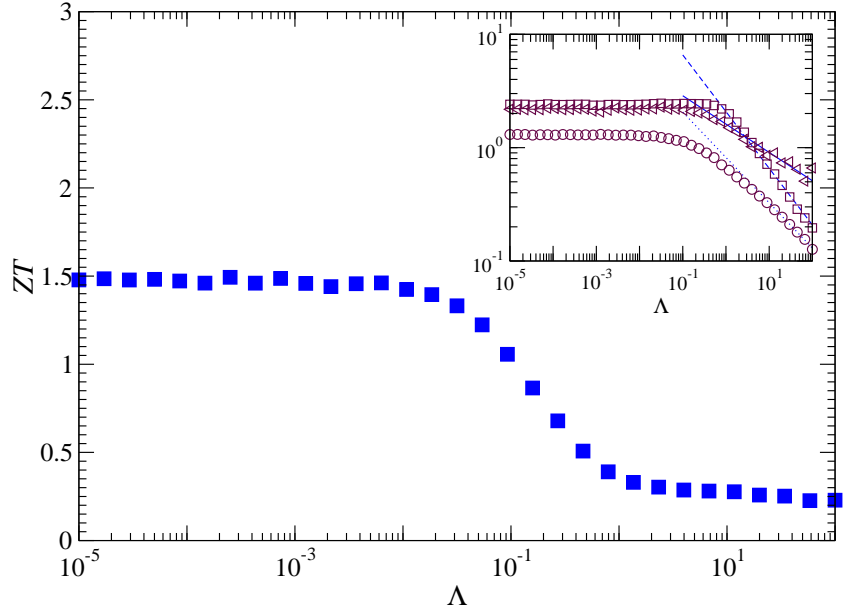}
\caption{Figure of merit $ZT$ (squares) as a function of the mass ratio
  $\Lambda$ for a monoatomic ideal gas ($D=2$).  The Onsager coefficients were
  obtained from two complementary experiments on a channel of $L=10$ cells,
  scatterers of radius $R=1$, and with parameters: {\bf Exp 1}) $T_\L=900$,
  $T_\R=1110$, $\mu_\L/T_\L = \mu_\R/T_\R = -6.3$, and {\bf Exp 2})
  $T_\L=T_\R=1000$, $\mu_\L/T_\L = -6.1$ and $\mu_\R/T_\R = -6.5$.  In the
  inset, we show the transport coefficients $\sigma$ (circles), $\kappa$
  (squares), and $S^2$ (triangles).  For the sake of presentation, $\sigma$
  has been scaled by a factor of $2000$.  The lines correspond to the scalings
  $\Lambda^{-1/4}$ (solid), $\Lambda^{-1/2}$ (dashed), and $\Lambda^{-2/5}$
  (dotted).
\label{fig:ZT}}
\end{figure}

To consider a gas of interacting particles we assume that the lattice disks of
mass $M$, freely rotate, so that when a particle collides with a disk, their
energies mixes, according to the collision rules \eref{eq:int-colrules}. This
induces a particle-particle interaction that is mediated by the collision with
the lattice disks.  The strength of the interaction is determined by the mass
ratio between disks and particles $\Lambda = M/m$ \cite{MMLL}.  The
noninteracting case is recover in the limit of $\Lambda \rightarrow 0$, when
the disks are much lighter than the particles.  Therefore, we can study the
effect of the interaction by tuning the value of $\Lambda$.  First we have
studied the TE transport for a gas of monoatomic particles ($D=2$).  In
Fig.~\ref{fig:ZT}, we show the TE figure-of-merit $ZT$ as a function of
$\Lambda$.  When $\Lambda\rightarrow0$, $ZT\rightarrow 3/2$, which is the
expected value of the noninteracting case (see Eq.~\ref{eq:ZTmain}).  In the
inset of Fig.~\ref{fig:ZT}, we also show the behaviour of the TE transport
coefficients.

Therefore, Eq.~\eref{eq:ZTmain} is an upper limit of the interacting case when
the interaction strength vanishes. This is behaviour is expected to be generic
since, as explained in \cite{CMMP08}, any interaction tends to correlate the
energy carried by the particles with the external gradients, thus decreasing
the efficiency.

\begin{figure}
  \includegraphics[height=.38\textheight]{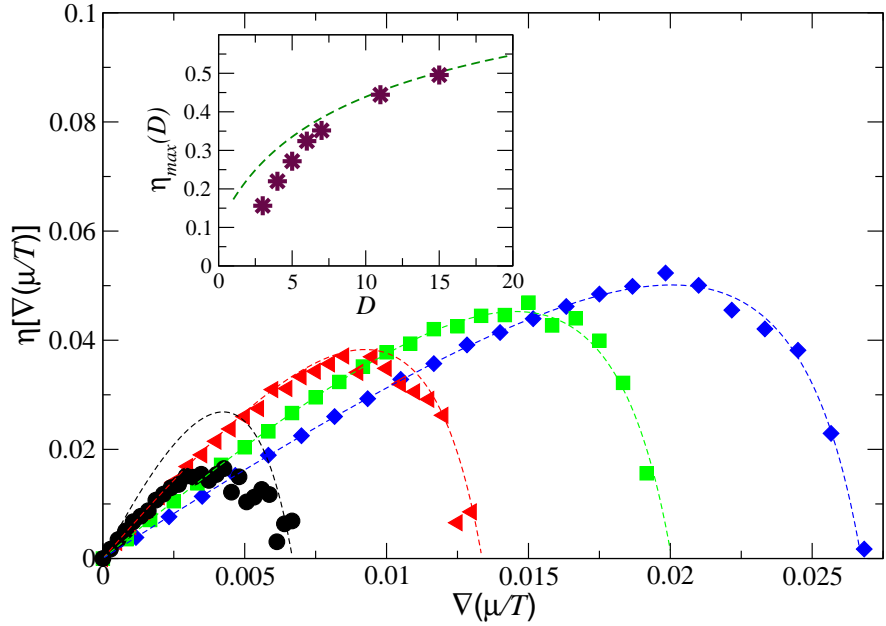}
  \caption{Thermodynamic    efficiency    $\eta$     as    a    function    of
  $\nabla\left(\mu/T\right)$  for  a chain  of  $15$  cells with  $\Lambda=1$,
  $T_L=95$,   $T_R=105$,  and   $D=3$  (circles),   $D=7$   (squares),  $D=11$
  (diamonds),  and $D=15$  (triangles).  The dashed  lines  correspond to  the
  analytical  solution obtained  from  substituting \eref{eq:noint-currents-2}
  into \eref{eq:efficiency-2}.  In the inset, maximum  efficiency obtained for
  the    different   values   of    $D$   (crosses),    in   units    of   the
  $\eta_\mathrm{carnot}$.  Here, the  results for $D=2$, $3$ and  $4$ are also
  shown.    The  dashed   line  corresponds   to  the   analytical  expression
  \eref{eq:efficiency}       for      a       gas       of      noninteracting
  particles.\label{fig:int-eta}}
\end{figure}

Next, we have studied the behaviour of the TE efficiency on $D$, for and gas
of interacting composite particles.  In Fig.~\ref{fig:int-eta}, we show the TE
efficiency $\eta$ as a function of $\nabla(\mu/T)$ for $\Lambda=1$, for
different values of $D$.  As for the noninteracting case, the efficiency
reaches a maximum value $\eta_{max}$ for some optimal value of the chemical
potential gradient.  However, consistently with the results of
Fig.~\ref{fig:ZT}, $\eta_{max}$ of the gas of interacting composite particles
is smaller than the noninteracting gas.  This can be seen for small $D$, in
the inset of Fig.~\ref{fig:int-eta} were $\eta_{max}$ is plotted for the
considered values of $D$.  For more complex particles (larger $D$),
$\eta_{max}$ approaches the noninteracting solution. Although the behaviour at
large $D$ is particular of the rotating-disk interaction, we cannot discard
that the same could be observed for a more generic type of
interactions. Indeed, when $D\gg 1$, the amount of energy of the lattice disk
becomes negligible compare to the total energy contained in the composite
particle.  Furthermore, for the gas of interacting particles, $\eta$ also
grows with $D$, indicating that the microscopic mechanism for the increase of
the TE efficiency discovered in \cite{CMMP08} also applies to the interacting
case.

\section{Conclusions}
\label{sec:concl}

We have studied the effect of interaction on the properties of thermoelectric
transport of gases of polyatomic molecules.  We have shown that while the TE
efficiency of gases of noninteracting molecules is an upper bound for the
efficiency of the interacting case when the interaction strength vanishes, the
later still increases with the complexity of the molecules.


\appendix

\section{Coupling with the thermo-chemical baths}
\label{app:baths}

Consider a infinite $2$-dimensional particle {\em reservoir} at equilibrium
temperature $T$ and density $\rho$, filled with point-like composite particles
of mass $m$ and $d_int$ \dof. The {\em reservoir}, coupled to the system
through an opening of section $\lambda$, exchanges particles with the system,
so that in the neighbourhood of the coupling the system is at equilibrium with
the {\em reservoir}.  Our aim is to obtain the rates at which particles and
energy are injected into the system.  Inside the {\em reservoir} each
component of the velocity of the particles is distributed according to the
Maxwell-Boltzmann distribution
\begin{equ} \label{eq:MB}
f_T(v) dv  =  2\pi  \left(\frac{m}{2\pi
  kT}\right)^{d/2} v^{d-1} e^{-\frac{mv^2}{2kT}}dv \ .
\end{equ}
Note that \eref{eq:MB} is independent of $d_{int}$.  To obtain the particle
injection rate $\gamma$ we need to compute how many particles of the {\em
reservoir} hit the opening per unit time.  From all the particles moving in a
given direction $\vec{v}$, the number of particles that cross the opening in
an infinitesimal time interval $dt$, are those contained in the parallelepid
of cross section $\lambda$ and length $v\cos(\theta)dt$, namely $\rho\lambda
v\cos(\theta)dt$.

Since the {\em reservoir} is at equilibrium, the probability that a particle
with speed between $v$ and $v+dv$ is injected into the system in a time
interval $dt$ is obtained as
\begin{equa}[2] \label{eq:Wv}
  &W(v)dvdt  & \  =&  \int_0^{2\pi}d\phi\int_0^{\pi/2}d\theta v^2sin\theta
  \left( \rho\lambda v\cos\theta\right) \left(2\pi\right)^{-1} f_T(v) dv dt \ , \\
  & & \ = \ & 2 \rho\lambda\left(\frac{m}{2\pi kT}\right) v^2
  e^{-\frac{mv^2}{2kT}}dvdt \ .
\end{equa}
Taking $v = \sqrt{2E/m}$ in \eref{eq:Wv}, where $E$ is the translational
energy, one obtains the probability that a particle with energy $E$ between
$E$ and $E+dE$ is injected into the system in a time interval $dt$ as
\begin{equ} \label{eq:WE}
W(E)dEdt  =   \frac{2\rho\lambda}{\pi^{1/2}(2\pi m kT)^{1/2}}
  \left(\frac{E}{kT}\right)^{1/2} e^{-\frac{E}{kT}}dEdt \ .
\end{equ}

In terms of $W(E)$, the particle injection rate is defined as $\gamma =
\int_0^\infty W(E)dE$, yielding
\begin{equ} \label{eq:gamma}
\gamma  =  \frac{\lambda}{\left(2\pi m\right)^{1/2}} \rho
\left(kT\right)^{1/2} \ .
\end{equ}
The energy injection rate is obtained as $\varepsilon = \gamma\langle
E\rangle$, where $\langle E\rangle$ is the mean energy of the injected
particles, namely $\langle E\rangle = \langle E_{trans}\rangle + \langle
E_{int}\rangle$, where $\langle E_{trans}\rangle$ is the mean energy of the
translational \dof~given by
\begin{equ}
\langle  E_{trans}\rangle = \frac{\int_0^\infty E W(E)dEdt}{\int_0^\infty
  W(E)dEdt} = \frac{3}{2}kT \ ,
\end{equ}
and $\langle E_{int}\rangle = \frac{d_{int}}{2}kT$ is the mean energy of the
internal \dof~Note that the mean translational energy of the injected
particles is not $\frac{d}{2}kT$ but $\frac{d+1}{2}kT$. Denoting the total
number of \dof~as $D=d+d_{int}$ we finally obtain
\begin{equ} \label{eq:epsilon}
\varepsilon  =  \frac{D+1}{2}\frac{\lambda}{\left(2\pi
    m\right)^{1/2}}\rho(kT)^{3/2} \ . 
\end{equ}

\section{Collision rules}
\label{app:colrules}

In this appendix we derive the collision rules that mix the energy among the
translational components of the particle's velocity, its internal degrees of
freedom.  We consider composite particles with $d_{int}$ internal rotational
\dof Each ``particle" of mass $m$ can be imagined as a stack of $d_{\rm int}$
small identical disks of mass $m/\D$ and radius $r \ll R$, rotating freely and
independently at a constant angular velocity $\omega_i$, $i=1,\ldots,
d_{int}$.  The center of mass of the particle moves with velocity
$\vec{v}=(v_x,v_y)$.

We assume that at a collision of a particle with a lattice disk, the normal
component of the particle's velocity is reflected, and the tangent component,
and internal angular velocities change, so that the total energy and local
momentum are conserved. Introducing the following notation
\begin{equa}[3] \label{eq:change}
& \xi_0 & \ = \ & v_t\\
& \xi_i & \ = \ & \alpha^{1/2}\omega_i \ , \ \ \ \textrm{for} \ i = 1, \ldots,
  {d_{int}} \ ,
\end{equa}
where $\alpha = \Theta/m$ and $\Theta = mr^2/2{d_{int}}$ is the moment of
inertia of the internal disks, the particle's energy becomes
\begin{equ} \label{eq:energy-xi}
E = \frac{m}{2}\left(v_n^2 + \sum_{i=0}^{d_{int}} \xi_i^2\right) \ .
\end{equ}
Denoting with primed (unprimed) variables the velocities after (before) the
collision, the equation for the energy conservation is
\begin{equ} \label{eq:E-const}
\sum_{i=0}^{d_{int}} {\xi'}_i^2 = \sum_{i=0}^{d_{int}} \xi_i^2 \ , 
\end{equ}
and for the conservation of local momentum
\begin{equ} \label{eq:L-const}
\xi'_0 + \alpha^{1/2}\sum_{i=1}^{d_{int}}{\xi'}_i = \xi_0 +
\alpha^{1/2}\sum_{i=1}^{d_{int}}\xi_i \ . 
\end{equ}
To solve for the primed momenta, we cast \eref{eq:L-const} as
\begin{equ}[3] \label{eq:torque}
\xi'_0 = \xi_0 - {d_{int}}\mathcal{K} \ \ \ \textrm{and} \ \
\xi'_i = \xi_i + \frac{1}{\alpha^{1/2}}\mathcal{K} \ .
\end{equ}
Substituting \eref{eq:torque} into \eref{eq:E-const} we obtain for the
collisional torque $\mathcal{K}$
\begin{equ} \label{eq:K}
\mathcal{K} = \frac{2\alpha}{{d_{int}}\left(1+{d_{int}}\alpha\right)}
\left({d_{int}}\xi_0 - \frac{1}{\alpha}\sum_{i=1}^{d_{int}} \xi_i\right)
\end{equ}
Substituting \eref{eq:K} back into \eref{eq:torque} and using \eref{eq:change}
the we obtain the collision rules
\begin{equa}[3] \label{eq:colrules}
& v_n' & \ = \ & -v_n \\
& v_t' & \ = \ & \frac{1-d_{int}\alpha}{1+d_{int}\alpha} v_t +
  \frac{2\alpha}{1+d_{int}\alpha} \sum_{i=1}^{d_{int}} \omega_i \\
& \omega_i' & \ = \ & \omega_i + \frac{2\alpha}{1+d_{int}\alpha} v_t -
  \frac{2}{d_{int}(1+d_{int}\alpha)} \sum_{i=1}^{d_{int}} \omega_i \ .
\end{equa}

For the interacting case, namely when the lattice disks rotate, these can be
considered as one additional rotor of mass $M$ and radius $R$ and angular
velocity $\overline{\omega}$. Following the same procedure, the reader can
easily verify that the collision rules are
\begin{equa}[3] \label{eq:int-colrules}
& v_n' & \ = \ & -v_n \\
& v_t' & \ = \ & (1-\Omega d_{int}^2) v_t + \Omega d_{int}\left(\overline{\omega} +
  \sum_{i=1}^{d_{int}} \omega_i\right) \\
& \overline{\omega}' & \ = \ & \overline{\omega} +
  \frac{\Omega}{R^2\Lambda}\left[d_{int}v_t - \left(\overline{\omega} +
  \sum_{i=1}^{d_{int}} \omega_i\right)\right] \\
& \omega_i' & \ = \ & \omega_i + \frac{\Omega}{\alpha}\left[d_{int}v_t - \left(\overline{\omega} +
  \sum_{i=1}^{d_{int}} \omega_i\right)\right] \ ,
\end{equa}
where $\Lambda=M/m$ and
\begin{equ} \label{eq:nu}
\Omega = 2\left(d_{int}^2 + \frac{d_{int}}{\alpha} + \frac{1}{R^2\Lambda}\right) \ .
\end{equ}

 These collision rules are a generalization of the ones introduced in
\cite{MMLL}. Thus, they are deterministic, time reversible and preserve the
energy and local angular momentum.


\begin{thebibliography}{1}
\bibitem{phonons} Assuming that particles do not interact means that we fully
  neglect the phonon transport.
\end{thebibliography}

\end{document}